\documentclass[journal]{IEEEtran}

\usepackage{amssymb}
\usepackage[ruled]{algorithm2e}
\usepackage{mathrsfs}
\usepackage{amsmath}
\usepackage{graphicx}
\usepackage{float}
\usepackage{subfigure}
\usepackage{amsfonts}
\usepackage{caption} 
\usepackage{color}
\usepackage{multirow}
\usepackage{bm}
\usepackage{stfloats}
\usepackage{amsthm}
\usepackage{color}
\usepackage{cite}



\newcommand{\rev}{\textcolor{black}}
\newcommand{\tcb}{\textcolor{black}}

\begin{document} 

\title{Message Passing Based Block Sparse Signal Recovery for DOA Estimation Using Large Arrays}   

\author{Yiwen Mao, Dawei Gao, Qinghua Guo, \IEEEmembership {Senior Member, IEEE} and Ming Jin
\thanks{Corresponding author: Q. Guo (qguo@uow.edu.au)}
\thanks{Y. Mao and Q. Guo are with the School of Electrical, Computer and Telecommunications Engineering, University of Wollongong, NSW, 2522, Australia (e-mail: maoyiwen1125@outlook.com; qguo@uow.edu.au).}
\thanks{D. Gao is with the Hangzhou Institute of Technology, Xidian University, Hangzhou 311200, China 
(e-mail: gaodawei@xidian.edu.cn).}
\thanks {M. Jin is with the Faculty of Electrical Engineering and Computer Science, Ningbo University, Ningbo 315211, China (e-mail:jinming@nbu.edu.cn). }
}

\maketitle

\begin{abstract} 
This work deals with directional of arrival (DOA) estimation with a large antenna array. We first develop \rev{a novel signal model} with a sparse system transfer matrix using an inverse discrete Fourier transform (DFT) operation, which leads to the formulation of a structured block sparse signal recovery problem with a sparse sensing matrix. This enables the development of a low complexity message passing based Bayesian \rev{algorithm with a factor graph representation}. Simulation results demonstrate the superior performance of the proposed method.  

\end{abstract}


\section{Introduction}
\IEEEPARstart {D}{irection-of-arrival (DOA)} estimation finds numerous applications in various areas such as wireless communications, acoustics and radar  \cite{kim1996two1}. In particular, the use of massive antenna arrays at base stations in  wireless communications \cite{Marzetta2010Noncooperative,Rusek2013Scaling} give rises to the problem of DOA estimation using large antenna arrays while with a limited number of snapshots (e.g., due to the movement of mobile terminals), \rev{which is the focus of this letter}.

Various DOA estimation methods have been proposed. The classical subspace based methods including the multiple signal classification (MUSIC) algorithm \cite{schmidt1986multiple3} and the estimation parameter via rotational invariance technique (ESPRIT) \cite{Roy1989ESPRIT4} can achieve high angular resolution,  which, however, have high computational complexity and require sufficient snapshots to obtain the subspace.  By exploiting the sparsity of source signals in the spatial domain, DOA estimation can be formulated as a compressive sensing (CS) problem and solved using sparse Bayesian learning (SBL) \cite{tipping2001sparse17, tipping2000The, wipf2004sparse6,ji2008bayesian16,Dai2016Root}. However,  conventional SBL based methods suffer from high computational complexity due to the matrix inversion involved, which is a concern in the case of large antenna arrays.
In the case of uniform linear array (ULA), the discrete Fourier transform (DFT) basis was used to construct a sparse representation for DOA estimation \cite{Rao2014Distributed,Chen2016Pilot,Shen2016Compressed}. However, the leakage of energy in the DFT basis can lead to significant 
performance loss \cite{Zhang2019Improved,Dai2018FDD}. 
A two stage DFT-based method was proposed in \cite{Cao2017A}, where coarse estimates of DOAs are found through DFT, and then the estimates are refined through search within local regions. \rev{However, difficulties can be experienced} in determining the number of sources at the stage of coarse estimation due to the energy leakage, and the refining may result in high complexity to achieve high precision. 


In this letter, exploiting the structure of the steering vector, we first develop a novel \rev{signal model} for DOA estimation with an inverse DFT operation. The model enables us to formulate the DOA estimation as a structured block sparse signal recovery problem. In particular, the sensing matrix is very sparse, which leads to a \rev{sparse factor graph} representation and facilities \rev{the development of a low complexity DOA estimation method with the message passing techniques} \tcb{\cite{Rangan2011Generalized,Krzakala2012Statistical,Huang2017Sparse,Dauwels2007On, bpa}}.
Simulation results are provided to demonstrate the advantages of the proposed method.  

\section{A Novel Sparse Modle for DOA Estimation} 
We assume a ULA with $M$ elements and half wavelength spacing. The steering vector with a DOA $\theta \in [-0.5{\pi},0.5{\pi}]$ can be represented as
\begin{equation}\label{g.1}
\begin{split}
  \mathbf{a}(\theta)&=[1, e^{-j \pi \sin(\theta)}, \ldots, e^{-j \pi (M-1) \sin(\theta)}]^{T} \\  
  &= [1, e^{-j\frac{2\pi}{M}(w_{\theta}+\alpha_{\theta})},\ldots,e^{-j\frac{2\pi}{M} (w_{\theta}+\alpha_{\theta}) (M-1)}]^{T} \\
  &= \mathbf{p}(w_{\theta}) \odot \mathbf{p}(\alpha_{\theta}) ,   
\end{split}
\end{equation}
\rev{where $ \mathbf{p}(c)=[1, e^{-j\frac{2\pi}{M} c}, \ldots, e^{-j\frac{2\pi}{M} (M-1) c }]^{T}$,  $\odot$ denotes \rev{the} element-wise product},  
\begin{equation}\label{g.2}
w_{\theta}+\alpha_{\theta}=0.5M \sin(\theta),
\end{equation}
with $w_{\theta}=\text{round}(0.5M \sin(\theta))$ being an integer in $[-0.5M, 0.5M]$ and $\alpha_{\theta}$ being a decimal in $[-0.5, 0.5)$, and round ($c$) returns an integer which is closest to $c$. Then we perform an $M$-point inverse DFT to  $\mathbf{a}(\theta)$, leading to 
\begin{eqnarray}\label{g.3}
\mathbf{d}(\theta)= \mathcal{F}^{-1} \mathbf{a}(\theta) = \mathbf {u}_{w_{\theta}}\circledast \mathcal{F}^{-1} \mathbf{p}(\alpha_{\theta}), 
\end{eqnarray}
where $\circledast$ denotes the circular convolution operation, $\mathbf {u}_{w_{\theta}} = \mathcal{F}^{-1} \mathbf {p}(w_{\theta})$ is a vector with only a single non-zero element located at the $[(w_{\theta}+M)_{\text{mod} M} +1]$th position. $\mathcal{F}^{-1} \mathbf{p}(\alpha_{\theta})$ is a vector with \rev{its} $l$th element given as 
\begin{equation}\label{g.4}
[\mathcal{F}^{-1} \mathbf{p}(\alpha_{\theta})]_l
= \tcb{\frac{1}{\sqrt{M}}\textstyle\sum_{m=0}^{M-1} e^{j\frac{2m\pi}{M}(l-1-\alpha_\theta)}}.
\end{equation}
According to (\ref{g.4}), we can find that $\mathcal{F}^{-1} (\mathbf{p}(\alpha_{\theta}))$ is a vector with only few significant elements at the beginning and end of the vector. Hence, with (\ref{g.3}), $\mathbf{d}(\theta)$ has few significant elements around the $[(w_{\theta}+M)_{\text{mod} M} +1  ]$th element. We then define a length-$L$ vector $\mathbf{g}_{\alpha_{\theta}}$ which consists of $L$ most significant elements in  $\mathcal{F}^{-1} \mathbf{p}(\alpha_{\theta})$, and ignore other elements. Then the circular convolution (\ref{g.3}) can be re-written in a matrix form as
\begin{equation}\label{g.5}
\mathbf{d}(\theta)= \mathbf{V}_{w_{\theta}}\mathbf{g}_{\alpha_{\theta}},
\end{equation}
where $\mathbf{V}_{w_{\theta}}$ has a size of $M \times L$, \rev{and it has} only a single non-zero element 1 in each column, and their locations correspond to the $L$ most significant elements in $\mathbf{d}(\theta)$.

Consider $K$ narrow-band far-field signals $\{s_k, k=1,2,\ldots, K\}$
impinging onto the ULA. Then the $M$-point inverse DFT of the received signal  can be represented as
\begin{eqnarray}\label{g.6}
 \mathbf{y} &=&  [\mathbf{d}(\theta_1), ..., \mathbf{d}(\theta_{K})] \mathbf{s} + \mathcal{F}^{-1} \mathbf{w}, \nonumber \\
 &=&[\mathbf{V}_{w_{\theta_1}},...,\mathbf{V}_{w_{\theta_K}}] \text{vec}([\mathbf{g}_{\alpha_{\theta_1}}s_1,...,\mathbf{g}_{\alpha_{\theta_K}}s_K]) + \mathbf{n},
\end{eqnarray}
where $\text{vec}(\cdot)$ \rev{denotes} the vectorization operation and $\mathbf{s}=[s_1,...,s_K]^T$, 
and $\mathbf{w}$ and $\mathbf{n}$ denote the noise vector and its inverse DFT, respectively. 
Once we obtain the non-zero elements locations in $\mathbf{V}_{w_{\theta_k}}$ and $\alpha_{\theta_k}$, according to (\ref{g.2}), we can get the DOA as $\theta_k=\text{arcsin} [2(w_{\theta_k}+\alpha_{\theta_k})/M]$.

However, (\ref{g.6}) is still difficult to use  as  the matrix $[\mathbf{V}_{w_{\theta_1}},...,\mathbf{V}_{w_{\theta_K}}]$ is unknown. To circumvent this problem, 
we extend the matrix $[\mathbf{V}_{w_{\theta_1}},...,\mathbf{V}_{w_{\theta_K}}]$ to $[\mathbf{V}_{w_{\theta^\prime_1}},...,\mathbf{V}_{w_{\theta^\prime_M}}]$ 
where $\{w_{\theta^\prime_m}, m = 1,2,\ldots,M\} = \{-0.5M, -0.5M+1, \ldots, 0.5M-1\}$. Hence, $\{{w_{\theta_k}},k=1,2,\ldots,K\}$ is a subset of $\{{w_{\theta^\prime_m}},m=1,2,\ldots,M\}$. Accordingly, we extend $\mathbf{s}=[s_1,...,s_K]^T$ to $\mathbf{s^\prime}=[s^\prime_1,...,s^\prime_M]^T$ by \rev{inserting} zeros to it. 
Then, we can construct the following sparse model
\begin{eqnarray}\label{g.7}
    \mathbf{y} &=& [\mathbf{V}_{w_{\theta^\prime_1}},...,\mathbf{V}_{w_{\theta^\prime_M}}] \text{vec}([\mathbf{g}_{\alpha_{\theta^\prime_1}}s^\prime_1,...,\mathbf{g}_{\alpha_{\theta^\prime_M}}s^\prime_M])  + \mathbf{n}  \nonumber \\
      &=&  \mathbf{V} \mathbf{x} + \mathbf{n},  
\end{eqnarray}
where the sensing matrix $\mathbf{V}$  is known and sparse (only a single non-zero element 1 in each column), and  $\mathbf{x}=\text{vec}([\mathbf{g}_{\alpha_{\theta^\prime_1}}s^\prime_1,...,\mathbf{g}_{\alpha^\prime_{\theta_M}}s^\prime_M])$ is \rev{a structured block sparse vector} with the elements of $\mathbf{g}_{\alpha_{\theta}}$ given in (\ref{g.4}). 

In the above, we only consider a single snapshot. If multiple snapshots are available, the received signal matrix can be represented as  
\begin{equation}\label{g.8}
   \mathbf{Y} =  \mathbf{V} \mathbf{X} + \mathbf{N},
\end{equation}
where $\mathbf{Y}=[\mathbf{y}(1),...,\mathbf{y}(T)]$, $\mathbf{X} = [\mathbf{x}(1),...,\mathbf{x}(T)]$ and $\mathbf{x}(t)= \text{vec}([\mathbf{g}_{\alpha_{\theta^\prime_1}}s^\prime_1(t),...,\mathbf{g}_{\alpha^\prime_{\theta_M}}s^\prime_M(t)])$.
Regarding the sparse model, we have the following remarks: 
\begin{itemize}
    \item 
    \rev{The new model does not have grid mismatch problem} and a DOA is jointly determined by $w_{\theta}$ and ${\alpha}_{\theta}$ (see (\ref{g.2})).  
    \item Our aim is to estimate the nonzero elements locations in $\mathbf{s^\prime}$ and the corresponding $\alpha_\theta$, i.e., recovering the block sparse vector $\mathbf{x}$. 
    \item Matrix $\mathbf{V}$ is \rev{highly} sparse, \rev{which leads to factor graph representation with long loops, facilitating the use of message passing techniques.}
    \item In the case of multiple snapshots, the common support can be exploited by estimating the vectors in $\mathbf{X}$ jointly.
\end{itemize}


\section{The Proposed Algorithm}

Inspired by the SBL \cite{tipping2001sparse17}, we use the sparsity inducing prior for  $\mathbf{s^\prime}(t)=[s^\prime_1(t),...,s^\prime_M(t)]^T$, and accommodate the common support of the vectors in $\mathbf{S^\prime}=[\mathbf{s^\prime}(1),...,\mathbf{s^\prime}(T)]$ through forcing them to share common precisions, i.e.,    
$p(\mathbf{S}^\prime | \boldsymbol{\gamma})=\textstyle\prod_{t=1}^{T} \textstyle\prod_{m=1}^{M} p(s^\prime_{m}(t) |\gamma_{m})$,
where $p(s^\prime_{m}(t) | \gamma_{m}) =\mathcal{C N}(s^\prime_{m}(t) ; 0, \gamma_{m}^{-1})$, $\bm{\gamma}=[\gamma_{1},...,\gamma_{m}]^T$ and $\mathcal{C N}(\cdot ;\boldsymbol{a}, \boldsymbol{C})$ denotes a complex Gaussian distribution with mean $\boldsymbol{a}$ and (co)variance $\boldsymbol{C}$. 
The precision $\gamma_{m}$ is Gamma distributed, denoted as $\Gamma(\gamma_{m} ; \epsilon, \eta)$.
We assume that the noise is zero mean complex Gaussian with precision $\lambda$, then
$p(\mathbf{Y} | \mathbf{X}, \lambda)=\textstyle\prod_{t=1}^{T} p(\mathbf{y}(t) | \mathbf{x}(t), \lambda)$,
where $p(\mathbf{y}(t) | \mathbf{x}(t), \lambda) = \mathcal{C N}(\mathbf{y}(t);\mathbf{V}\mathbf{x}(t),\lambda^{-1}\mathbf{I})$. 
The prior of $\alpha_m$ is a uniform distribution over $[-0.5,0.5)$, i.e., $p{(\alpha_m)}=U[-0.5,0.5)$. So the joint  distribution of the variables can be factorized as 
\begin{equation}\label{5.100}
\begin{aligned}
&p( \lambda,\mathbf{X},\mathbf{S}^\prime, \mathbf{G},\bm{\alpha}, \bm{\gamma} | \mathbf{Y}) \\
& \propto p(\mathbf{Y}|\mathbf{X},\lambda) p(\mathbf{X}|\mathbf{S}^\prime,\mathbf{G}) p(\mathbf{S}^\prime|\bm{\gamma}) p(\bm{\gamma}) p(\mathbf{G}|\bm{\alpha})  p(\bm{\alpha}) p(\lambda) \\
& \propto f_{\lambda} \textstyle\prod_{t} f_{\mathbf{y}(t)} \textstyle\prod_{m,l} f_{x_{m,l}(t)} f_{s^\prime_m(t)} f_{\gamma_m} f_{g_{l,m}} f_{\alpha_m},
\end{aligned}
\end{equation}
where $\mathbf{G}=[\mathbf{g}_{\alpha_{\theta^\prime_1}},...,\mathbf{g}_{\alpha_{\theta^\prime_M}}]$, $\bm{\alpha}=[\alpha_1,...,\alpha_M]^T$, and the factors and corresponding functions are listed in Table I. 
\begin{table}[hbp]
\centering
\caption{Factors and corresponding distributions in (\ref{5.100}).}
\begin{tabular}{ccc}
\hline
Factor & Distribution & Function\\
\hline
$f_{\lambda}$ & $p(\lambda)$ & $\lambda^{-1}$ \\

$f_{\mathbf{y}(t)}$ & $p\left(\mathbf{y}(t) | \mathbf{x}(t), \lambda\right)$ & $\mathcal{C N}(\mathbf{y}(t);\mathbf{V}\mathbf{x}(t),\lambda^{-1}\mathbf{I})$ \\

$f_{x_{m,l}(t)}$ &$p(x_{m,l}(t) | s^\prime_m(t),g_{l,m})$ & $\delta\left(x_{m,l}(t) - s^\prime_m(t) \times g_{l,m}\right)$ \\

$f_{{s^\prime_m(t)}}$ & $p(s^\prime_m(t) |\gamma_{m})$ & $\mathcal{C N}(s^\prime_m(t) ; 0, \gamma_{m}^{-1})$\\

$f_{\gamma_{m}}$ & $p(\gamma_{m})$ & $\Gamma\left(\gamma_{m} ; \epsilon, \eta\right)$\\

\tcb{$f_{g_{l,m}}$} & \hspace{-1.2cm}\tcb{$p(g_{l,m} | \alpha_m)$} & \hspace{-1.4cm}\tcb{$\delta(g_{l,m}-\frac{1}{\sqrt{M}}\sum_{m'=0}^{M-1} e^{j\frac{2m'\pi}{M}(l-1-\alpha_m)})$} \\

$f_{\alpha_m}$ & $p(\alpha_m)$ & $U[-0.5,0.5)$\\

\hline
 \end{tabular}
 \label{t1}
\end{table}

\begin{figure}[t]
	\centering
\includegraphics[scale=0.5]{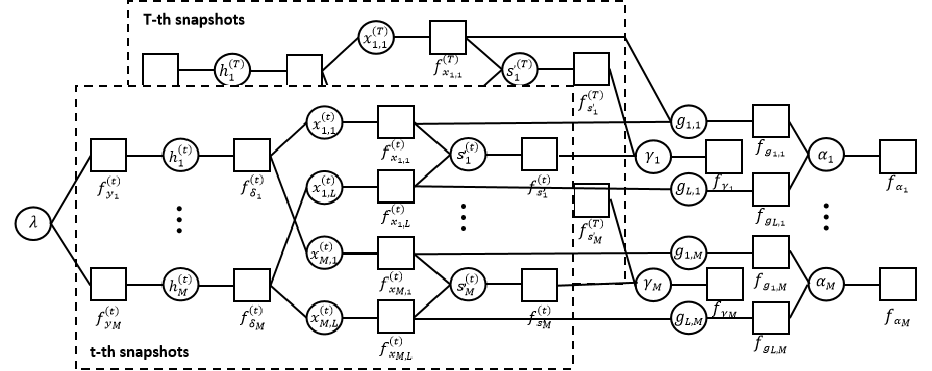}
	\caption{Graph representation \rev{of} (\ref{5.100}).}
	\label{figfactor_graph}
\end{figure}

Our aim is to obtain the marginals $\{p(\gamma_m| \mathbf{Y})\}$ and $\{p(\alpha_{m}| \mathbf{Y})\}$, so that their a posteriori estimates can be obtained, based which the DOAs \tcb{can be extracted}. However, it is difficult to \tcb{compute} the exact posteriors. We use message passing based approximate inference techniques based on a factor graph representation of (\ref{5.100}) 
\rev{in Fig.~\ref{figfactor_graph}}. 
 It is noted that the left degree of variable node $x_{m,l}(t)$ is $L$ and the right degree of function node $f_{\delta_{m}(t)}$ is only $1$ due to the special structure of $\mathbf{V}$ (\rev{see Fig.~\ref{figfactor_graph}}, $L=2$ as a toy example). \rev{The graph has large loops, which facilitates the design of message passing algorithms with good convergence and low complexity.} Next, we derive a message passing algorithm, \rev{based on the mean field (MF) \cite{Dauwels2007On} and  belief propagation (BP) \cite{bpa}. The algorithm is an iterative one with a forward recursion and a backward recursion in each iteration, where 
 the message computations in current round iteration will use some messages computed in the last round. We use arrows to denote the direction of message passing.} 
 
\subsection{Forward Recursion}
With belief $b(\lambda)$ computed in (\ref{5.65}) (in the last round iteration), the forward message $\mathcal{M}_{f_{y_{m}}\rightarrow h_{m}}$ \rev{can be computed by MF, i.e.,
$\mathcal{M}_{f_{y_{m}}\rightarrow h_{m}} \propto \exp \{\langle \log f_{y_{m}} \rangle_{b(\lambda)} \} 
\propto \mathcal{C} \mathcal{N}(h_{m} ; y_{m}, \hat{\lambda}^{-1})$,
where $\hat{\lambda}=\langle\lambda\rangle_{b(\lambda)}$ is the estimate of $\lambda$. The message $\tcb{\mathcal{M}_{f_{\delta_{m}}\rightarrow x_{m,l}}}$ 
is computed by the BP, yielding}
$\mathcal{M}_{f_{\delta_{m}}\rightarrow x_{m,l}} = \langle f_{\delta_{m}} \rangle_{ \mathcal{M}_{h_{m}\rightarrow f_{\delta_{m}}}\prod_{m^{'},l^{'}\neq m,l} \mathcal{M}_{x_{m^{'},l^{'}}\rightarrow f_{\delta_{m}}}} 
\propto \mathcal{C} \mathcal{N}(x_{m,l};\overrightarrow{x}_{m,l},\overrightarrow{v}_{x_{m,l}})$,
where
\begin{equation}\label{5.8}
\overrightarrow{x}_{m,l} = y_{m} - \hat{p}_m + \overleftarrow{x}_{m,l},
\end{equation}
\begin{equation}\label{5.9}
\overrightarrow{v}_{x_{m,l}} = \hat{\lambda}^{-1} + v_{p_{m}} - \overleftarrow{v}_{x_{m,l}},
\end{equation} 
\begin{equation}\label{5.10}
\tcb{\hat{p}_m = \textstyle\sum\overleftarrow{x}_{m,l}, ~ v_{p_{m}} = \textstyle\sum \overleftarrow{v}_{x_{m,l}}.}
\end{equation}
We note that (\ref{5.10}) 
involve\tcb{s} only $L$ additions. With the incoming message $\mathcal{M}_{x_{m,l}\rightarrow f_{x_{m,l}}} = \mathcal{M}_{f_{\delta_{m}}\rightarrow x_{m,l}}$ and the factor $f_{x_{m,l}} = \delta(x_{m,l} - s^\prime_m g_{l,m})$, we can obtain an intermediate function node $\tilde{f}_{x_{m,l}}$ \rev{according to BP,} i.e.,
$\tilde{f}_{x_{m,l}} = \textstyle\int f_{x_{m,l}}\mathcal{M}_{x_{m,l}\rightarrow f_{x_{m,l}}}{\tcb{dx_{m,l}}} 
\propto \mathcal{C} \mathcal{N}(s^\prime_m g_{l,m};\overrightarrow{x}_{m,l},\overrightarrow{v}_{x_{m,l}})$.
\tcb{Then we can compute the message $\mathcal{M}_{f_{x_{m,l}}\rightarrow s^\prime_m} \propto \mathcal{C} \mathcal{N}(s^\prime_m;\overrightarrow{s}^\prime_{m,l},\overrightarrow{v}_{s^\prime_{m,l}})$,
where}
\begin{equation}\label{5.14}
\overrightarrow{s}^\prime_{m,l} = \overrightarrow{x}_{m,l}/\hat{g}_{l,m},
\end{equation}
\begin{equation}\label{5.15}
\overrightarrow{v}_{s^\prime_{m,l}} = \overrightarrow{v}_{x_{m,l}}/|\hat{g}_{l,m}|^2.
\end{equation}
$\hat{g}_{l,m}$ is the mean of the complex Gaussian belief of $g_{l,m}$, which is computed in (\ref{5.54}). The message $\tcb{\mathcal{M}_{s^\prime_m\rightarrow f_{s^\prime_m}}}$ 
\tcb{is} the product of all the complex Gaussian message $\mathcal{M}_{f_{x_{m,l}}\rightarrow s^\prime_m}$, i.e.,
$\tcb{\mathcal{M}_{s^\prime_m\rightarrow f_{s^\prime_m}}} = \textstyle\prod_l \mathcal{M}_{f_{x_{m,l}}\rightarrow s^\prime_m} 
\propto \mathcal{C} \mathcal{N} (s^\prime_m;\overrightarrow{f}_{s^\prime_m},\overrightarrow{v}_{f_{s^\prime_m}})$,
where
\begin{equation}\label{5.17} 
\overrightarrow{v}_{f_{s^\prime_m}} = (\textstyle\sum_l 1/\overrightarrow{v}_{s^\prime_{m,l}})^{-1},
\end{equation}
\begin{equation}\label{5.18}
\overrightarrow{f}_{s^\prime_m} = \overrightarrow{v}_{f_{s^\prime_m}}(\textstyle\sum_l \overrightarrow{s}^\prime_{m,l}/\overrightarrow{v}_{s^\prime_{m,l}}).
\end{equation}
\rev{With message $\tcb{\mathcal{M}_{f_{s^\prime_m}\rightarrow s^\prime_m}} \propto \mathcal{C} \mathcal{N}(s^\prime_m;0,\hat{\gamma}_m^{-1})$ and $\hat{\gamma}_m$ later computed in (\ref{5.41}), the belief $b(s^\prime_m)$ can be obtained as} 
$b(s^\prime_m) \propto \tcb{\mathcal{M}_{s^\prime_m\rightarrow f_{s^\prime_m}}} \tcb{\mathcal{M}_{f_{s^\prime_m}\rightarrow s^\prime_m}} 
= \mathcal{C} \mathcal{N}(s^\prime_m;\hat{s}^\prime_m,\hat{v}_{s^\prime_m})$,
where
\begin{equation}\label{5.20}
\hat{s}^\prime_m = \overrightarrow{f}_{s^\prime_m}/(1+\overrightarrow{v}_{f_{s^\prime_m}}\hat{\gamma}_m), 
\end{equation}
\begin{equation}\label{5.21}
v_{s^\prime_m} = (1/\overrightarrow{v}_{f_{s^\prime_m}} + \hat{\gamma}_m)^{-1}.
\end{equation}
\tcb{According to MF,}
$\tcb{\mathcal{M}_{f_{s^\prime_m}\rightarrow \gamma_m}} \propto \exp \{\langle \log f_{s^\prime_m} \rangle_{b(s^\prime_m}) \}  \propto \gamma_m\exp \{ -\gamma_m(|s^\prime_m|^2 + \overrightarrow{v}_{f_{s^\prime_m}})\}$,
so that the belief $b(\gamma_m)$ of the hyperparameter $\gamma_m$ is given as
$b(\gamma_m)  \propto f_{\gamma_m}\textstyle\prod_t \tcb{\mathcal{M}_{f_{s^\prime_m}(t)\rightarrow \gamma_m}}
\propto \gamma_m^{\epsilon+T} \exp \{ -\gamma_m (\eta + \textstyle\sum_{t} (|s^\prime_m(t)|^2 + \overrightarrow{v}_{f_{s^\prime_m}}(t)))\}$.
The message from $\mathcal{M}_{f_{x_{m,l}}}$ to $g_{l,m}$ is computed \rev{with the MF rule,} i.e.,
$\mathcal{M}_{f_{x_{m,l}}\rightarrow g_{l,m}} = \exp\{\langle \log \tilde{f}_{x_{m,l}} \rangle_{b(s^\prime_m}) \} \propto \mathcal{C} \mathcal{N} (g_{l,m};\overrightarrow{g}_{l,m},\overrightarrow{v}_{g_{l,m}})$,
where
\begin{equation}\label{5.25}
\overrightarrow{g}_{l,m} =\overrightarrow{x}_{m,l}(s^\prime_m)^{*}/(|\hat{s}^\prime_m|^2 + v_{s^\prime_m}),
\end{equation}
\begin{equation}\label{5.26}
\overrightarrow{v}_{g_{l,m}} = \overrightarrow{v}_{x_{m,l}}/(|\hat{s}^\prime_m|^2 + v_{s^\prime_m}).
\end{equation}   
Then the message $\mathcal{M}_{g_{l,m}\rightarrow f_{g_{l,m}}}$ \tcb{is} the product of all the Gaussian messages $\mathcal{M}_{f_{x_{m,l}(t)}\rightarrow g_{l,m}}$, i.e.,
$\mathcal{M}_{g_{l,m}\rightarrow f_{g_{l,m}}} = \textstyle\prod_t \mathcal{M}_{f_{x_{m,l}(t)}\rightarrow g_{l,m}} 
\propto \mathcal{C} \mathcal{N} (g_{l,m};\overrightarrow{f}_{g_{l,m}},\overrightarrow{f}_{v_{g_{l,m}}})$,
where
\begin{equation}\label{5.28}
\overrightarrow{v}_{f_{g_{l,m}}} = (\textstyle\sum_t 1/\overrightarrow{v}_{g_{l,m}}(t))^{-1},
\end{equation}
\begin{equation}\label{5.29}
\overrightarrow{f}_{g_{l,m}} = \overrightarrow{v}_{f_{g_{l,m}}}(\textstyle\sum_t \overrightarrow{g}_{l,m}(t)/\overrightarrow{v}_{g_{l,m}}(t)).
\end{equation}
\tcb{Note} that $f_{g_{l,m}}$ is a nonlinear function, \tcb{making} the computation of the message $\mathcal{M}_{f_{g_{l,m}}\rightarrow \alpha_m}$ difficult. To solve this problem, $f_{g_{l,m}}$ is linearized \tcb{with} the first order Taylor expansion, i.e.,
$f_{g_{l,m}}(\alpha_m) \approx f_{g_{l,m}}(\hat{\alpha}_m^{'}) + f_{g_{l,m}}^{'}(\hat{\alpha}_m^{'})(\alpha_m - \hat{\alpha}_m^{'})$,
with
\begin{equation}\label{5.31}
f_{g_{l,m}}^{'}(\hat{\alpha}_m^{'}) = M^{-\frac{1}{2}}\textstyle\sum_{m'=0}^{M-1} -j\frac{2m'\pi}{M}e^{j\frac{2m'\pi}{M}(l-1-\hat{\alpha}_m^{'})}, \nonumber
\end{equation}
where $\hat{\alpha}_m^{'}$ denotes the estimate of $\alpha_m$ in the last iteration. 
\tcb{The message $\mathcal{M}_{f_{g_{l,m}}\rightarrow \alpha_m} = \mathcal{N}(\alpha_m;\overrightarrow{\alpha}_{m,l},\overrightarrow{v}_{{\alpha}_{m,l}})$, where
$\overrightarrow{v}_{{\alpha}_{m,l}} = (1/\overrightarrow{v}_{{\alpha}_{m,l}}^R + 1/\overrightarrow{v}_{{\alpha}_{m,l}}^I)^{-1}$,
$\overrightarrow{\alpha}_{m,l} = \overrightarrow{v}_{{\alpha}_{m,l}}(\overrightarrow{\alpha}_{m,l}^R/\overrightarrow{v}_{{\alpha}_{m,l}}^R + \overrightarrow{\alpha}_{m,l}^I/\overrightarrow{v}_{{\alpha}_{m,l}}^I)$,
\begin{equation}\label{5.35}
\overrightarrow{\alpha}_{m,l}^R \!= \!\frac{\Re\{\!\!\overrightarrow{f}_{g_{l,m}}\!\}\!-\!\Re\{f_{g_{l,m}}(\hat{\alpha}_m^{'})\}\!+\!\Re\{f_{g_{l,m}}^{'}(\hat{\alpha}_m^{'})\}\hat{\alpha}_m^{'}}{\Re\{f_{g_{l,m}}^{'}(\hat{\alpha}_m^{'})\}}
\end{equation}
\begin{equation}\label{5.36}
\overrightarrow{v}_{{\alpha}_{m,l}}^R = \frac{1}{2}\frac{\overrightarrow{v}_{f_{g_{l,m}}}}{|\Re\{f_{g_{l,m}}^{'}(\hat{\alpha}_m^{'})\}|^2},
\end{equation}
and $\overrightarrow{\alpha}_{m,l}^I$ and $\overrightarrow{v}_{{\alpha}_{m,l}}^I$ have the same expressions 
except replacing $\Re\{\cdot\}$ with $\Im\{\cdot\}$.}

Note that $\alpha_m$ has a uniform prior over  $[-0.5,0.5)$. We simplify the computation of the belief $b(\alpha_m)$ as   
$b\left(\alpha_m\right) = \mathcal{N}\left(\alpha_m, \hat{\alpha}_m, v_{\alpha_m}\right) $
with
\begin{equation}\label{5.38}
v_{\alpha_m} = (\textstyle\sum_l 1/\overrightarrow{v}_{{\alpha}_{m,l}})^{-1},
\end{equation}
\begin{equation}\label{5.39}
\hat{\alpha}_m = v_{\alpha_m}(\textstyle\sum_l \overrightarrow{\alpha}_{m,l}/\overrightarrow{v}_{{\alpha}_{m,l}}),
\end{equation}
when $\hat{\alpha}_m \in (-0.5, 0.5)$. Otherwise, it is clipped to 0.5 or -0.5.

\subsection{Backward Recursion}
\rev{The message $\tcb{\mathcal{M}_{f_{s^\prime_m}\rightarrow s^\prime_m}}$ is computed by MF,} i.e.,
$\mathcal{M}_{f_{s^\prime_m}\rightarrow s^\prime_m} \propto \exp \{\langle \log f_{s^\prime_m}\rangle_{b(\gamma_m)} \} 
\propto \mathcal{C} \mathcal{N} (s^\prime_m;0,\hat{\gamma}_m^{-1})$,
where
\begin{equation}\label{5.41}
\hat{\gamma}_m = (\epsilon+T)/[\eta + \textstyle\sum_t (|s^\prime_m(t)|^2 + \overrightarrow{v}_{f_{s^\prime_m(t)}})].
\end{equation}
\tcb{Then, we use the rule in \cite{Luo2021Unitary} to automatically tune $\epsilon$ as}
\begin{equation}\label{6.3}
\tcb{\epsilon = 0.5 \textstyle\sqrt{\log(\frac{1}{M}\textstyle\sum_{m} \hat{\gamma}_m) - \frac{1}{M}\textstyle\sum_{m}\log \hat{\gamma}_m}.}
\end{equation}
The message $\tcb{\mathcal{M}_{s^\prime_m\rightarrow f_{x_{m,l}}}}$ is updated by the BP rule, i.e.,
$\tcb{\mathcal{M}_{s^\prime_m\rightarrow f_{x_{m,l}}}} = b(s^\prime_m)/\mathcal{M}_{f_{x_{m,l}}\rightarrow s^\prime_m} 
\propto \mathcal{C} \mathcal{N} (s^\prime_m;\tcb{\overleftarrow{f}_{s^\prime_m \rightarrow x_{m,l}}},\overleftarrow{v}_{f_{s^\prime_m} \rightarrow x_{m,l}})$,
where
\begin{equation}\label{5.43}
\overleftarrow{v}_{f_{s^\prime_m} \rightarrow x_{m,l}} = ( 1/v_{s^\prime_m} - 1/\overrightarrow{v}_{s^\prime_{m,l}} )^{-1},
\end{equation}
\begin{equation}\label{5.44}
\overleftarrow{f}_{s^\prime_m \rightarrow x_{m,l}} = \overleftarrow{v}_{f_{s^\prime_m} \rightarrow x_{m,l}} ( \hat{s}^\prime_m/v_{s^\prime_m} -\overrightarrow{s}^\prime_{m,l}/\overrightarrow{v}_{s^\prime_{m,l}}).
\end{equation}
The backward message $\mathcal{M}_{\alpha_m \rightarrow f_{g_{l,m}}}$ is Gaussian with mean $\overleftarrow{f}_{g_{l,m}}$ and variance $\overleftarrow{v}_{f_{g_{l,m}}}$, which can be calculated as
\begin{equation}\label{5.45}
\overleftarrow{v}_{f_{g_{l,m}}} = ( 1/v_{\alpha_m} - 1/\overrightarrow{v}_{{\alpha}_{m,l}} )^{-1},
\end{equation}
\begin{equation}\label{5.46}
\overleftarrow{f}_{g_{l,m}} = \overleftarrow{v}_{f_{g_{l,m}}} (\hat{\alpha}_m/v_{\alpha_m} -\overrightarrow{\alpha}_{m,l}/\overrightarrow{v}_{{\alpha}_{m,l}}).
\end{equation}
Then the message $\mathcal{M}_{f_{g_{l,m}}\rightarrow g_{l,m}}$ is \rev{calculated by the BP rule,}
$\mathcal{M}_{f_{g_{l,m}}\rightarrow g_{l,m}} = \textstyle\int f_{g_{l,m}}\mathcal{M}_{\alpha_m \rightarrow f_{g_{l,m}}}d \alpha_m 
\propto \mathcal{C} \mathcal{N} (g_{l,m};\overleftarrow{g}_{l,m},\overleftarrow{v}_{g_{l,m}})$,
where
\begin{equation}\label{5.50}
\overleftarrow{g}_{l,m} = f_{g_{l,m}}(\hat{\alpha}_m^{'}) + f_{g_{l,m}}^{'}(\hat{\alpha}_m^{'})(\overleftarrow{f}_{g_{l,m}} - \hat{\alpha}_m^{'}),
\end{equation}
\begin{equation}\label{5.51}
\overleftarrow{v}_{g_{l,m}} = \overleftarrow{v}_{f_{g_{l,m}}}|f_{g_{l,m}}^{'}(\hat{\alpha}_m^{'})|^2.
\end{equation}

Hence the belief $b(g_{l,m})$ is obtained as 
$b(g_{l,m}) \propto \mathcal{M}_{g_{l,m}\rightarrow f_{g_{l,m}}}\mathcal{M}_{f_{g_{l,m}}\rightarrow g_{l,m}}
= \mathcal{C} \mathcal{N} (g_{l,m};\hat{g}_{l,m},v_{g_{l,m}})$,
where
\setlength\abovedisplayskip{3pt}
\belowdisplayskip=5pt
\begin{equation}\label{5.53}
v_{g_{l,m}} = ( 1/\overrightarrow{v}_{f_{g_{l,m}}} + 1/\overleftarrow{v}_{g_{l,m}} )^{-1},
\end{equation}
\begin{equation}\label{5.54}
\hat{g}_{l,m} = v_{g_{l,m}} ( \overrightarrow{f}_{g_{l,m}}/\overrightarrow{v}_{f_{g_{l,m}}} + \overleftarrow{g}_{l,m}/\overleftarrow{v}_{g_{l,m}}).
\end{equation}
Then the message $\mathcal{M}_{g_{l,m}\rightarrow f_{x_{m,l}}}$ is updated by the BP rule, i.e.,
$\mathcal{M}_{g_{l,m}\rightarrow f_{x_{m,l}}} = b(g_{l,m})/\mathcal{M}_{f_{x_{m,l}}\rightarrow g_{l,m}} \propto \mathcal{C} \mathcal{N} (g_{l,m};\overleftarrow{f}_{g_{l,m} \rightarrow x_{m,l}},\overleftarrow{v}_{f_{g_{l,m} \rightarrow x_{m,l}}})$
where
\begin{equation}\label{5.56}
\overleftarrow{v}_{f_{g_{l,m} \rightarrow x_{m,l}}} = ( 1/v_{g_{l,m}} - 1/\overrightarrow{v}_{g_{l,m}} )^{-1},
\end{equation}
\begin{equation}\label{5.57}
\overleftarrow{f}_{g_{l,m} \rightarrow x_{m,l}} = \overleftarrow{v}_{f_{g_{l,m} \rightarrow x_{m,l}}} ( \hat{g}_{l,m}/v_{g_{l,m}} \!\!-\!\!\overrightarrow{g}_{l,m}/\overrightarrow{v}_{g_{l,m}} ).
\end{equation}
By combing the incoming messages $\tcb{\mathcal{M}_{s^\prime_m\rightarrow f_{x_{m,l}}}}$ and $\mathcal{M}_{g_{l,m}\rightarrow f_{x_{m,l}}}$, the message $\mathcal{M}_{f_{x_{m,l}}\rightarrow x_{m,l}}$ is complex Gaussian with mean $\overleftarrow{x}_{m,l}$ and variance $\overleftarrow{v}_{x_{m,l}}$, given as
\begin{equation}\label{5.58}
\overleftarrow{x}_{m,l} = \overleftarrow{f}_{s^\prime_m \rightarrow x_{m,l}}\overleftarrow{f}_{g_{l,m} \rightarrow x_{m,l}},
\end{equation}
\begin{equation}\label{5.59}
\begin{split}
&~~~~~~~~~~~~~~\overleftarrow{v}_{x_{m,l}}  = |\overleftarrow{f}_{s^\prime_m \rightarrow x_{m,l}}|^2\overleftarrow{v}_{f_{g_{l,m} \rightarrow x_{m,l}}} \tcb{+}
 \\
& |\overleftarrow{f}_{g_{l,m} \rightarrow x_{m,l}}|^2 \overleftarrow{v}_{f_{s^\prime_m} \rightarrow x_{m,l}}+ \overleftarrow{v}_{f_{g_{l,m} \rightarrow x_{m,l}}} \overleftarrow{v}_{f_{s^\prime_m} \rightarrow x_{m,l}}.
\end{split}
\end{equation}      
The message $\mathcal{M}_{f_{\delta_m}\rightarrow h_m}$ is \rev{updated by BP,} i.e.,
$\mathcal{M}_{f_{\delta_m}\rightarrow h_m} = \langle f_{\delta_{m}} \rangle_{\prod_{m^{'},l^{'}} \mathcal{M}_{x_{m^{'},l^{'}}\rightarrow f_{\delta_{m}}}} 
 \propto \mathcal{C} \mathcal{N} (h_m;\hat{p}_m,v_{p_{m}})$.
Then we compute the belief 
$b(h_m) = \mathcal{M}_{h_{m}\rightarrow f_{\delta_{m}}}\mathcal{M}_{f_{\delta_m}\rightarrow h_m} 
\propto \mathcal{C} \mathcal{N} (h_m;\hat{h}_m,\hat{v}_{h_m})$
where
\begin{equation}\label{5.62}
\setlength\abovedisplayskip{3pt}
\hat{v}_{h_m} = (\hat{\lambda} + 1/v_{p_{m}})^{-1},
\end{equation}
\begin{equation}\label{5.63}
\hat{h}_m = \hat{v}_{h_m}(y_m\hat{\lambda} + \hat{p}_m/v_{p_{m}}).
\setlength\belowdisplayskip{1pt}
\end{equation}      
The message $\mathcal{M}_{y_m\rightarrow \lambda}$ can be calculated by the MF rule, i.e.,
$\mathcal{M}_{y_m\rightarrow \lambda} = \textstyle\exp \{\langle \log f_{y_{m}}\rangle_{b(h_m)} \} 
\propto \lambda \exp\{-\lambda[|y_{m} - \hat{h}_m|^2 + \hat{v}_{h_m}] \}$.
With the \tcb{prior of $\lambda$,}  
the belief $b(\lambda)$ is updated by
\begin{equation}\label{5.65}
\begin{aligned}
b(\lambda) & 
\propto f_{\lambda} \textstyle\prod_{m,t} \mathcal{M}_{y_m(t)\rightarrow \lambda} \\
& \propto \lambda^{MT-1}\exp \{-\lambda \textstyle\sum_{m,t}[|y_{m}(t) - \hat{h}_m(t)|^2 + \hat{v}_{h_m(t)}]\}.
\end{aligned}
\end{equation}
Then the parameter $\lambda$ can be estimated as
\begin{equation}\label{5.66}
\hat{\lambda} = MT/\textstyle\sum_{m,t}[|y_{m}(t) - \hat{h}_m(t)|^2 + \hat{v}_{h_m(t)}].
\end{equation}
\tcb{The proposed algorithm is summarized in Algorithm 1.}

\begin{algorithm}[t]\label{algo1}
    \DontPrintSemicolon        
   \rev{ \caption{Proposed DOA Estimation Algorithm}        
    Initialization: $\epsilon = 0.01$, $\eta = 0.0001$, $\hat{\lambda} = 1$, $\overleftarrow{x}_{m,l}=0$, $\overleftarrow{v}_{x_{m,l}}=1$, $\hat{g}_{l,m}=1$, $\hat{s}^\prime_m=0$, $v_{s^\prime_m}=1$, and $\hat{\alpha}_m^{0}=0$.\;
    \SetKwRepeat{Do}{do}{while}
    \textbf{Repeat:}
    \\
    Update $\hat{\gamma}_m$ by (\ref{5.8}) $\sim$ (\ref{5.18}), (\ref{5.41}) and (\ref{6.3}) \\
    Update $\hat{s}^\prime_m$ and $v_{s^\prime_m}$ by (\ref{5.20}) and (\ref{5.21}) \\
    Update $\hat{\alpha}_m$ by (\ref{5.25}) $\sim$ (\ref{5.39}) \\
    Update $\hat{g}_{l,m}$ and $v_{g_{l,m}}$ by  (\ref{5.43}) $\sim$ (\ref{5.54}) \\
    Update $\hat{\lambda}$ by (\ref{5.56}) $\sim$ (\ref{5.66}) \\
\textbf{While} $\frac{\sum_{t}\Arrowvert \hat{s}^\prime_m(t)^{n+1} - \hat{s}^\prime_m(t)^{n} \Arrowvert_2^2} {\sum_{t}\Arrowvert  \hat{s}^\prime_m(t)^{n} \Arrowvert_2^2} > \sigma_s$. \\
Output the estimates of angles based on  $\{\hat{\gamma}_m, \hat{\alpha}_m$\}.}
\end{algorithm}


 


\begin{figure}[t]
    \centering
    \includegraphics[scale=0.4]{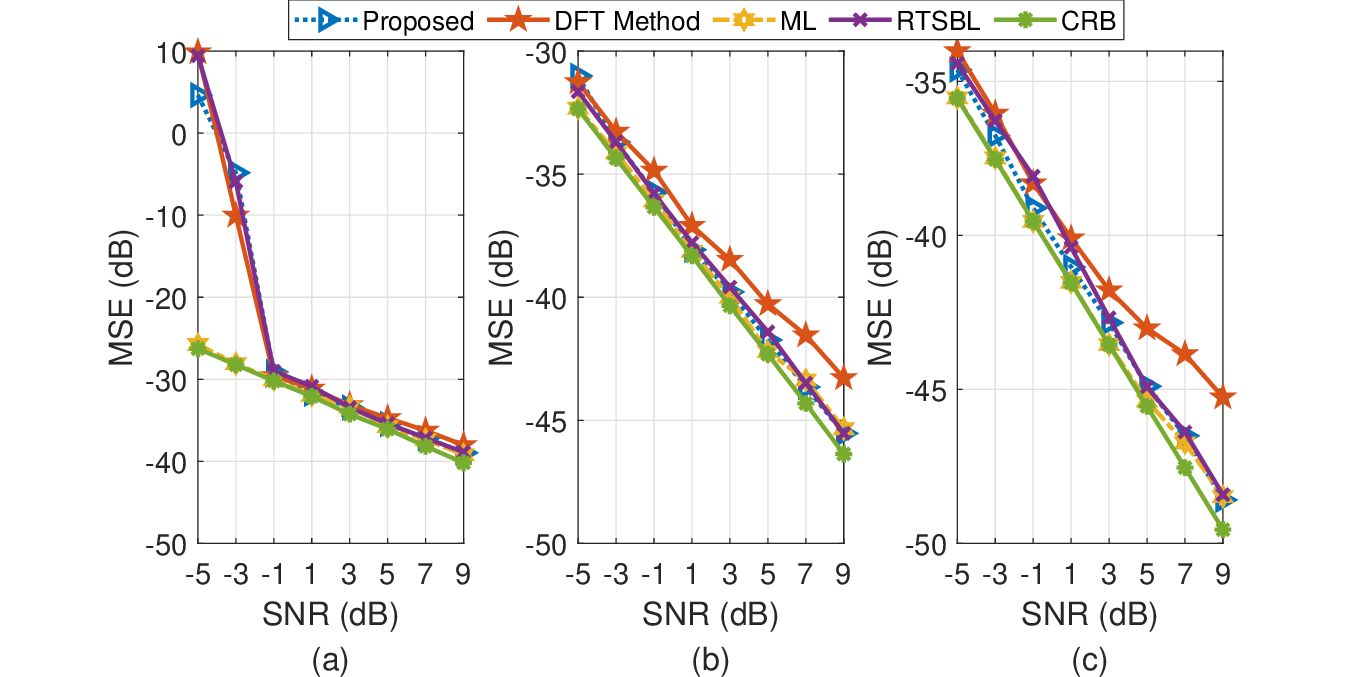}
    \caption{\tcb{Performance Comparison: $T=3(a), 10(b), 20(c)$.}}
    \label{FigSNR}
\end{figure}

\subsection{Complexity Analysis}

The proposed method needs to perform inverse fast Fourier transform (IFFT) as pre-processing with complexity $\mathcal{O}(MT\log M)$. For the message passing algorithm, the complexity is $\mathcal{O}(LMT)$ per iteration. As a comparison, the complexity of the DFT-based DOA method in \cite{Cao2017A} is $\mathcal{O}(MT\log M + M + PKMT)$, where $P$ determines the precision in fine search involved in the method and is typically a large number to obtain a high precision. The complexity of the root-SBL method (RTSBL) in \cite{Dai2016Root} is $\mathcal{O}(M^{3} + M^{2}T)$ per iteration. As $L \ll P$ and $L \ll M$, the complexity of our method can be significantly lower than these methods.

\section{Simulation Results}
We compare the proposed method with the DFT-based method  \cite{Cao2017A}, the RTSBL method \cite{Dai2016Root} and the maximum likelihood (ML) method \cite{stoica1989music5}. The Cramer–Rao bound (CRB) is also shown for reference.  
We assume that the number of array elements $M=128$, and choose $L=7$. The number of sources $K = 3$, and their corresponding DOAs are randomly drawn from the angle intervals $[-60^{\circ}, -50^{\circ}]$, $[-20^{\circ}, -10^{\circ}]$ and $[20^{\circ}, 30^{\circ}]$, respectively. 





\begin{figure}[t]
    \centering
    \includegraphics[scale=0.4]{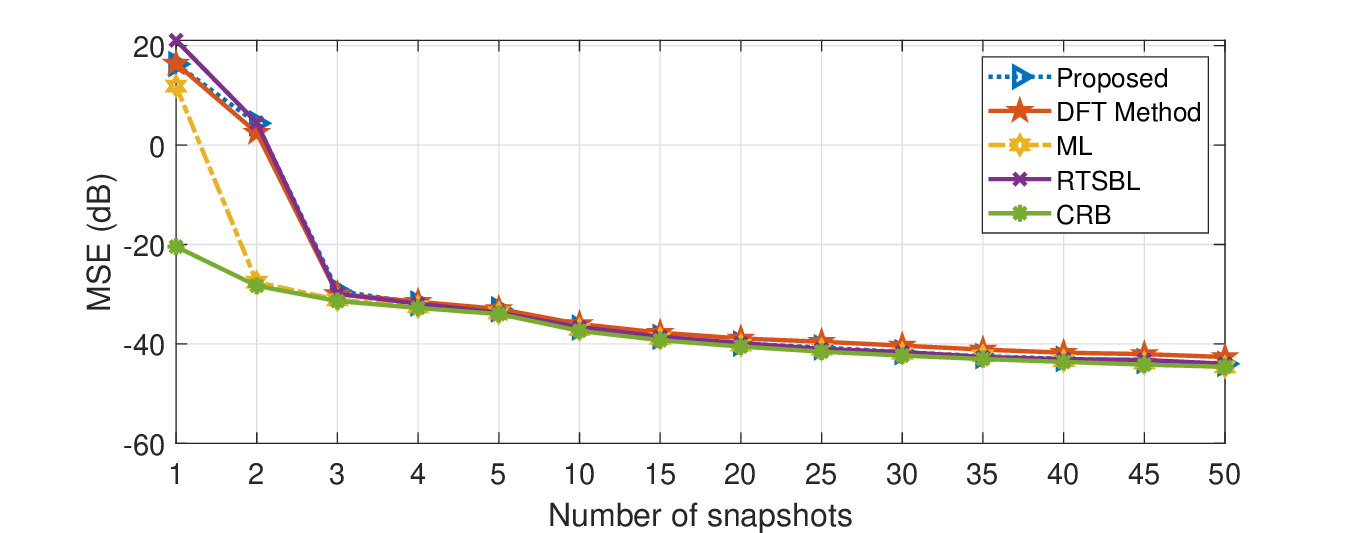}
    \caption{\tcb{Performance of the methods versus snapshot number.}}
    \label{FigSnapshots}
\end{figure}

\tcb{Figure~\ref{FigSNR} shows the mean squared error (MSE) performance of the methods versus SNR, where the number of snapshots $T = 3,10$ and $20$, respectively. It can be seen that the proposed method outperforms the DFT-based method, and it delivers performance similar to that of RTSBL, while with much lower complexity. The ML method performs slightly better but with extremely high complexity due to the exhaustive search.} 

The impact of the snapshot number on the performance of the methods is shown in Fig.~\ref{FigSnapshots}, where SNR=$0$dB, and the number of snapshots $T$ is varied from 1 to 50. We have similar observations as in Fig.~\ref{FigSNR}, and the proposed method performs well while with low complexity. 
From the above results, we can see that the proposed method is a promising alternative in the case of large arrays.




\clearpage
\bibliographystyle{IEEEtran}
\bibliography{Reference}

\end{document}